# Preventing SARS-CoV-2 superspreading events with antiviral intranasal sprays

George Booth*[1,2], Christoforos Hadjichrysanthou[1,3], Keira L Rice[2], Jacopo Frallicciardi[2], Zoltán Magyarics[2], Frank de Wolf[1], Jaap Goudsmit[2,4], Anna L Beukenhorst[†2,5], Roy Anderson[†1]

[1] Department of Infectious Disease Epidemiology, School of Public Health, Imperial College London, London, UK
[2] Leyden Laboratories BV, Leiden, The Netherlands
[3] Department of Mathematics, University of Sussex, Brighton, UK
[4] Departments of Epidemiology, Immunology and Infectious Diseases, Harvard TH Chan School of Public Health, Boston, MA, USA
[5] Department of Biostatistics, Harvard TH Chan School of Public Health, Boston, MA, USA

[†]These authors contributed equally to this work.

*Corresponding author: George Booth

**Email:** george.booth@leydenlabs.com

**Abstract**

Superspreading events are known to disproportionally contribute to onwards transmission of epidemic and pandemic viruses. Preventing infections at a small number of high-transmission settings is therefore an attractive public health goal. Here, we use deterministic and stochastic mathematical modelling to quantify the impact of intranasal sprays in containing outbreaks at a known superspreading event (the 2020 SARS-CoV-2 outbreak at the Diamond Princess cruise ship) and a conference event that led to extensive transmission.

We find that in the Diamond Princess cruise ship case study, there exists a 7–14-day window of opportunity for widespread prophylactic spray usage to significantly impact the number of infections averted. Given an immediate response to a known SARS-CoV-2 outbreak, alongside testing and social distancing measures, prophylactic efficacy and coverage greater than 65% could reduce the average number of infections by over 90%. In the conference case study, in the absence of additional public health interventions, analyses suggest much higher prophylactic efficacies and coverages are required to achieve a similar outcome. However, prophylactic use can half an individual's probability of being infected, and significantly reduce the probability of developing a severe infection. These results suggest that at a known potential superspreading event, early use of intranasal sprays can complement quarantining measures and significantly suppress a SARS-CoV-2 outbreak, even at suboptimal coverage. At a *potential* superspreading event of short duration, intranasal sprays can reduce individuals' risk of infection, but cannot prevent all infections or onwards community transmission.

**Introduction**

The probability distribution of the number of secondary cases generated by individuals infected by respiratory viruses is known to be highly skewed. The variance in the number of secondary cases is much bigger than the mean; a few individuals contribute disproportionally to onward transmission[1,2]. Such "superspreading events" have occurred in all virus outbreaks, especially for respiratory viruses capable of airborne transmission[1,2]. For example, in March 2020, a conference held in Boston resulted in the infection with SARS-CoV-2 in 99 of 175 attendees. Subsequent genetic analysis determined that nearly 2% of all reported cases in the US in 2020 could be traced back to this single event[3]. Following a religious conference in Kuala Lumpur, Malaysia, 600 attendees who tested positive for SARS-CoV-2 were found to be connected to nearly 50% of all infections in the country two months later, as well as to outbreaks in seven other countries[4].

To mitigate epidemics and pandemics, targeting superspreading events presents an attractive solution: preventing onward infection at a small number of events could greatly reduce or slow viral spread in the wider population. Existing interventions for preventing viral spread are not ideal to prevent superspreading events. Vaccines against respiratory viruses, like SARS-CoV-2 in particular, typically protect against disease, but not always against infection and onward transmission[5,6]. During the SARS-CoV-2 pandemic, non-pharmaceutical interventions prevented superspreading events by limiting close contacts and discouraging attendance at events involving many people. These measures were effective[2,7,8], but strongly relied on individual compliance. They were also costly to enforce and unsustainable to retain permanently[7]. Social distancing measures, where implemented robustly, disrupted society, education, and routine healthcare.

To mitigate super-spreading events and curb the spread of respiratory viruses like SARS-CoV-2, there is a need for innovative interventions that can be seamlessly integrated into healthcare protocols. An innovative intervention that could play a role in preventing superspreading events is an antibody-based intranasal spray. Various intranasal sprays with SARS-CoV-2 antibodies have been developed and tested for efficacy in clinical trials often involving a large number of participants (numbers in the range of 556 to 1336)[9–12]. These sprays demonstrated fast-acting protective efficacy against SARS-CoV-2 in a pre-exposure setting, as well as up to 72 hours post-exposure. Moreover, intranasal sprays have been suggested to prevent infection and onward transmission by neutralizing the virus at the port of entry[10].

A better understanding of the potential role intranasal sprays could play in preventing outbreaks of respiratory viruses is clearly required in preparations for the control of future epidemics. To prevent all onwards transmission using intranasal sprays, a prophylactic with high, continued efficacy is required, plus high coverage in the target population[13]. Focusing on superspreading events could provide an exceptionally efficient public health measure: preventing an important part of onward community transmission by intervening at a small number of events[1]. However, it is unknown if intranasal sprays could achieve this, and under what conditions regarding efficacy, coverage, compliance, and timing of use.

In this paper, we use deterministic and individual-based stochastic mathematical models to quantify the impact of intranasal sprays in containing outbreaks at two superspreading events for SARS-CoV-2. The first is the 2020 Diamond Princess cruise ship outbreak among passengers and crew, based on actual data of symptomatic infections after testing[14]. The second is a potential superspreading event at the 2009 French Society for Hospital Hygiene (SFHH) conference in Nice, France, with 405 attendees[15,16]. In this example, we use the recorded contacts to simulate an outbreak of SARS-CoV-2. We focus on quantifying the potential impact of intranasal sprays in containing outbreaks at these events. We examine scenarios of varying uptake, efficacy, and adherence to optimise guidelines that could be used to effectively contain viral spread.

**Methods**

The classic susceptible-exposed-infectious-recovered (SEIR) mathematical model was extended to describe COVID-19 disease transmission in two distinct, closed population scenarios (Figure 1).

Scenario I considers the real-life transmission of SARS-CoV-2 aboard the Diamond Princess Cruise ship, which departed from Yokohama, Japan on the 20th January 2020 with 2666 passengers and 1045 crew members on board (median age of passengers: 69, median age of crew members: 36[14]). Following the first confirmed COVID case on 1st February, symptomatic testing commenced on the 3rd February, and by the 5th February, all passengers were quarantined to their cabins. By the end of the cruise on 21st February, 314 confirmed symptomatic cases had been reported. The nature of this scenario constitutes a strictly closed population, with known infected individuals being removed from the population. To model this scenario, we adapt the mathematical model developed in Emery et al. (2020), which is stratified by passengers and crew members[14]. We parameterise this model via literature recorded estimates of key epidemiological features (i.e. incubation, infectious and generation time periods), as well as fitting our model to data describing known infections over time in passenger and crew populations on board the Diamond Princess via Bayesian inference methods[14]. A schematic diagram of the model is presented in Figure 2, with further details provided in the Supporting Information.

In Scenario II, we simulate a hypothetical SARS-CoV-2 outbreak during a five-day conference. We use contact data from the 2009 SPHH conference in Nice, France, consisting of 405 participants[15,16]. We develop an individual-based stochastic simulation utilising the Gillespie algorithm[17] to simulate the order and timing of disease transmission events more accurately. The interactions of participants are described by a contact network developed from real face-to-face interactions between participants, with elements of the contact matrix weighted relative to the frequency of interactions. Treatment compliance is related to treatment regimen via Claxton et al. (2001), who investigated the relationship between self-administered treatment frequency and adherence[18,19]. It is assumed that the treatment is administered for the full duration of the conference, and we track the progression of infections in the original population resulting from the conference for 45 days following its end. The schematic diagram of the model is presented in Figure 4, with further details provided in the Supporting Information.

In each scenario, we model the population-level transmission of SARS-CoV-2 in the presence and absence of a fast-acting prophylactic treatment. In both cases, it is assumed that the virus is novel, and individuals do not have pre-existing immunity. To tackle virus spread, we assume that the prophylactic treatment is immediately available to a fixed proportion of the population from the beginning of the simulation and acts to reduce the rate of virus transmission proportionally to the efficacy of the treatment. In Scenario II, where we differentiate between mild and severe infections, we also assume that the treatment reduces the probability of developing a severe infection proportionally to the treatment efficacy.

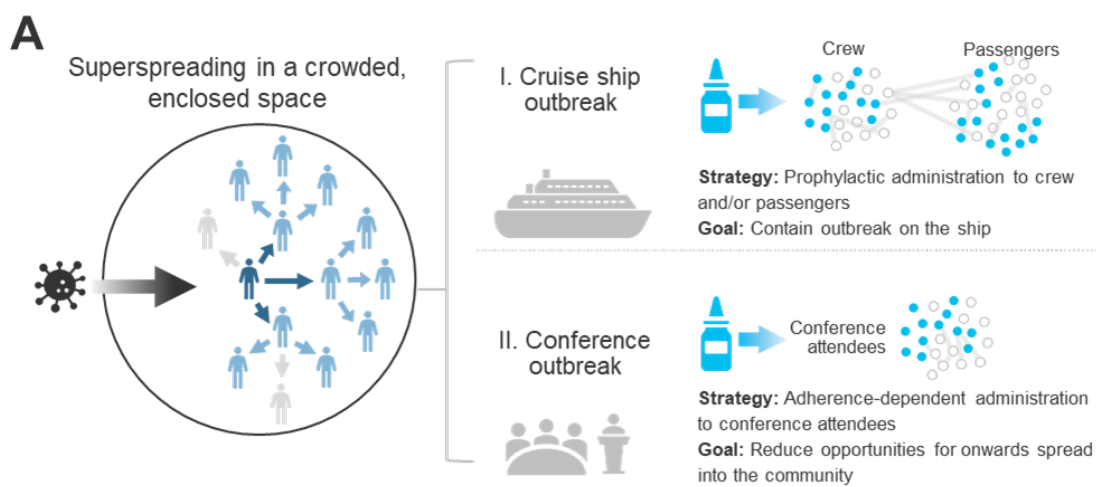

**Figure 1. A)** Schematic of the modelling framework. We consider two scenarios that capture the potential impact of intranasal prophylaxis on superspreading in closed populations: 1) Superspreading on board a cruise ship: a real-life example of a past outbreak where the goal for mitigation is to contain the outbreak on the ship and prevent onward transmission from the ship; and 2) superspreading at a conference: a simulated outbreak where the goal for pandemic mitigation is to reduce opportunities for onwards spread into the community. **B)** Table showing the features of the two scenarios.

**Results**

*Known superspreading event*

In the Diamond Cruise ship scenario (Figure 2), there is a known virus outbreak occurring over a one-month timescale, with supplementary public health interventions such as symptomatic testing (from 3rd February) and quarantine measures (from 5th February) also in place.

Figure 3A shows that minimising the time between discovery of the first case and administering intranasal sprays is crucial to averting the maximum number of infections (assuming 100% population adherence and 70% efficacy). Intervening within the first 7 days following the first known infection prevents on average 87% (intervention on day 0) to 100% (intervention on day 7) of total infections. When intervening in the second week (day 7 – day 14), secondary infections have had sufficient time to incubate, resulting in larger variability in intervention outcome between simulations. The secondary infections lead to an exponential increase in the number of productively infectious individuals, but a sufficiently large percentage of the population remain susceptible to infection, and therefore protectable by intranasal sprays. Intervening beyond day 21 from detection of the first case does not reduce the number of infections, given that most of the population have already been infected.

Simulations also show that in addition to intervention time, intervention efficacy and coverage strongly influence population-level outcomes. To avert an average of 80% of infections, we would only need to cover around 35% of the population (treatment from day 0, 100% efficacy; Fig 3B). In case of 100% coverage, a treatment efficacy of only 35% achieves the same result. Isolines in Figure 3B identify the range of combinations of intervention efficacy and coverage required to achieve a prescribed outcome. Interestingly, we see that under combinations of moderate efficacy and coverage (e.g. > 65%), more than 90% of infections can be averted, suggesting that during the onset of a known pandemic, intranasal sprays could reduce infections in a known superspreading event. To eliminate *any* onwards transmission altogether (bringing the basic reproduction number ($R_0$) below 1 by averting >99.9% of infections), intranasal sprays would need to reach ~95% efficacy and coverage, as shown in the Supporting Information. Therefore, despite the closed nature of this population, it remains difficult to completely prevent onward transmission once the SARS-CoV-2 outbreak has taken off.

We then modelled the effect of suboptimal coverage of intranasal sprays, in two scenarios: optimal coverage of the people at highest risk of severe COVID-19 only, or optimal coverage of people at occupational risk of COVID-19 only.

Coverage of 100% of the closed population could be more realistic for the elderly passengers than for the younger crew. During the COVID-19 pandemic, high-risk groups were more likely to comply with interventions due to fears and uncertainties surrounding infection cause and outcome[21]. The crew consisted of younger individuals at reduced risk of severe disease, and less likely to comply with protective treatments[21]. If coverage of the crew is suboptimal, it is not possible to avert all infections even at 100% efficacy (Figure 3C).

During the pandemic, some occupational groups were more likely to comply with protective treatments and interventions. High compliance was either caused by workplace orders of compliance, or because of higher occupational risk of infection with SARS-CoV-2. If coverage of the crew is optimal, but coverage of the passengers is not, a significantly smaller number of infections are averted due to the relatively smaller number of crew compared to passengers (Figure 3C).

**Table 1.** Summary of model parameters for Scenario I (Diamond Princess cruise ship). Infection parameters are either taken directly from or fitted to data presented in Emery et al. (2020)[14].

| PARAMETER | DESCRIPTION | VALUE |
|---|---|---|
| **INFECTION** | | |
| $N_p$ | Total number of passengers | 2666 |
| $N_c$ | Total number of crew | 1045 |
| $\bar{\beta}$ | Baseline transmission rate | |
| $c_{pp}$ | Probability of contact between passengers ($p$) | |
| $c_{cp} = c_{pc} = 0.1 c_{pp}$ | Probability of contact between passengers ($p$) and crew ($c$) | |
| $b_1$ | Transmission rate logistic parameter | Fitted |
| $\tau_{pp} = \tau_{cp} = \tau_{pc}$ | Transmission time delay between passengers ($p$) and crew ($c$) | |
| $\tau_{cc}$ | Transmission time delay between crew ($c$) | |
| $\theta_a$ | Transmission correction factor for asymptomatic infections | |
| $\theta_p$ | Transmission correction factor for pre-symptomatic infections | |
| $\chi$ | Probability of asymptomatic infection | |
| $N_{tests}$ | Number of COVID tests per day | Presented in Emery et al. (2020) [14] |
| $b_2$ | Transmission rate logistic parameter | 80 |
| $\nu$ | Transition rate from exposed to infectious | 2/4.3 |
| $\gamma_a$ | Rate of recovery from asymptomatic infection | 1/5 days$^{-1}$ |
| $\gamma_p$ | Rate of transition from pre-symptomatic to symptomatic | 1/2.1 days$^{-1}$ |
| $\gamma_s$ | Rate of recovery from symptomatic infection | 1/2.9 days$^{-1}$ |
| $m$ | Rate of removal from the ship | $\begin{cases} 0, & 0 \leq t < 17 \\ 0.7, & t \geq 17 \end{cases}$ |
| $h$ | Rate of transition from testing positive to testing negative | 1/7 days$^{-1}$ |
| $f$ | Fraction of symptomatic cases with known onset date | 199/314 |
| $c_{cc}$ | Probability of contact between crew ($c$) | 1 |
| $a_1$ | Transmission time delay parameter | 0.3 |
| $\alpha_2$ | Transmission parameter correction factor | -0.43 |
| **TREATMENT** | | |
| $\varepsilon$ | Efficacy | 70% |
| $q_p$ | Passenger adherence to treatment | 50% |
| $q_c$ | Crew adherence to treatment | 50% |

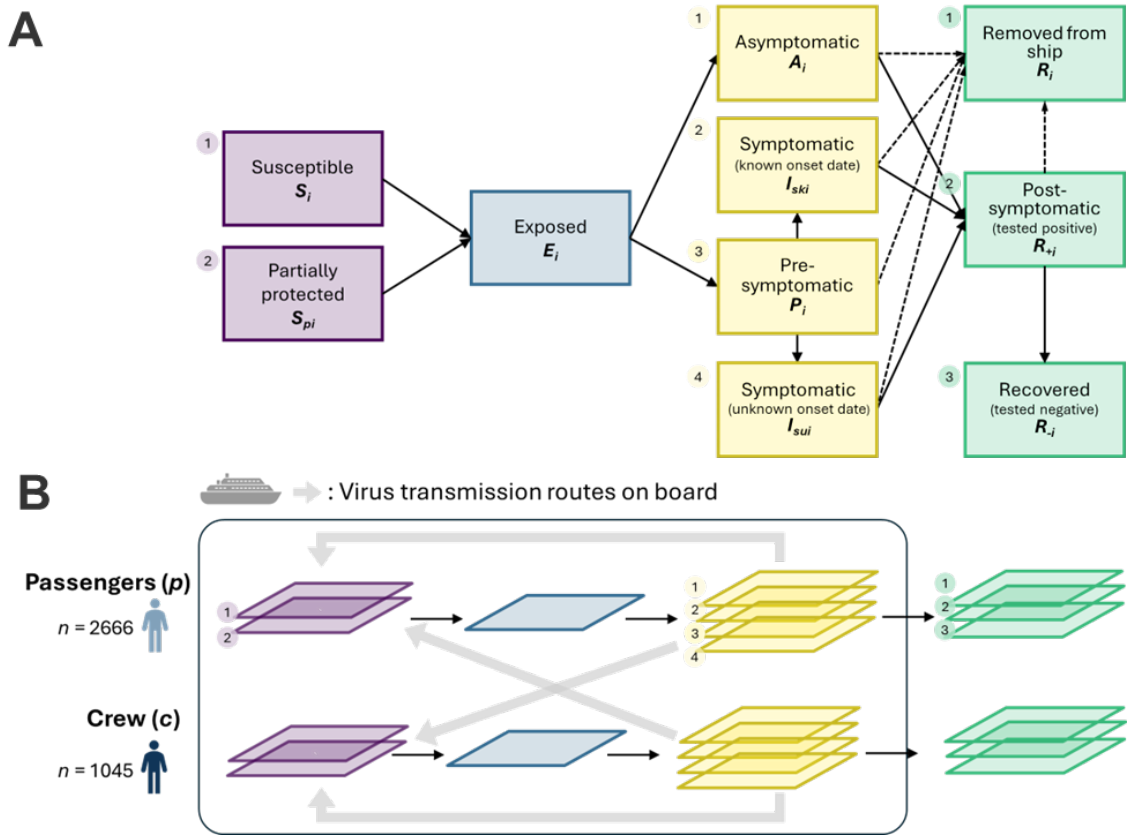

**Figure 2.** Schematic of the SEIR model for the Diamond Princess cruise ship. **A)** Adapted SEIR model for each subpopulation $i = p,c$. Susceptible passengers ($p$) or crew ($c$) can be either be unprotected or partially protected with an intranasal prophylactic. Susceptible individuals become exposed to the virus, which develops to either symptomatic or asymptomatic infection. Infected individuals then recover, and/or are removed from the ship. **B)** Schematic of virus transmission routes within and between subpopulations. For each subpopulation, we have a separate adapted SEIR model to describe infection dynamics. Infectious passengers can infect susceptible passengers and crew, and infectious crew can infect susceptible passengers and crew.

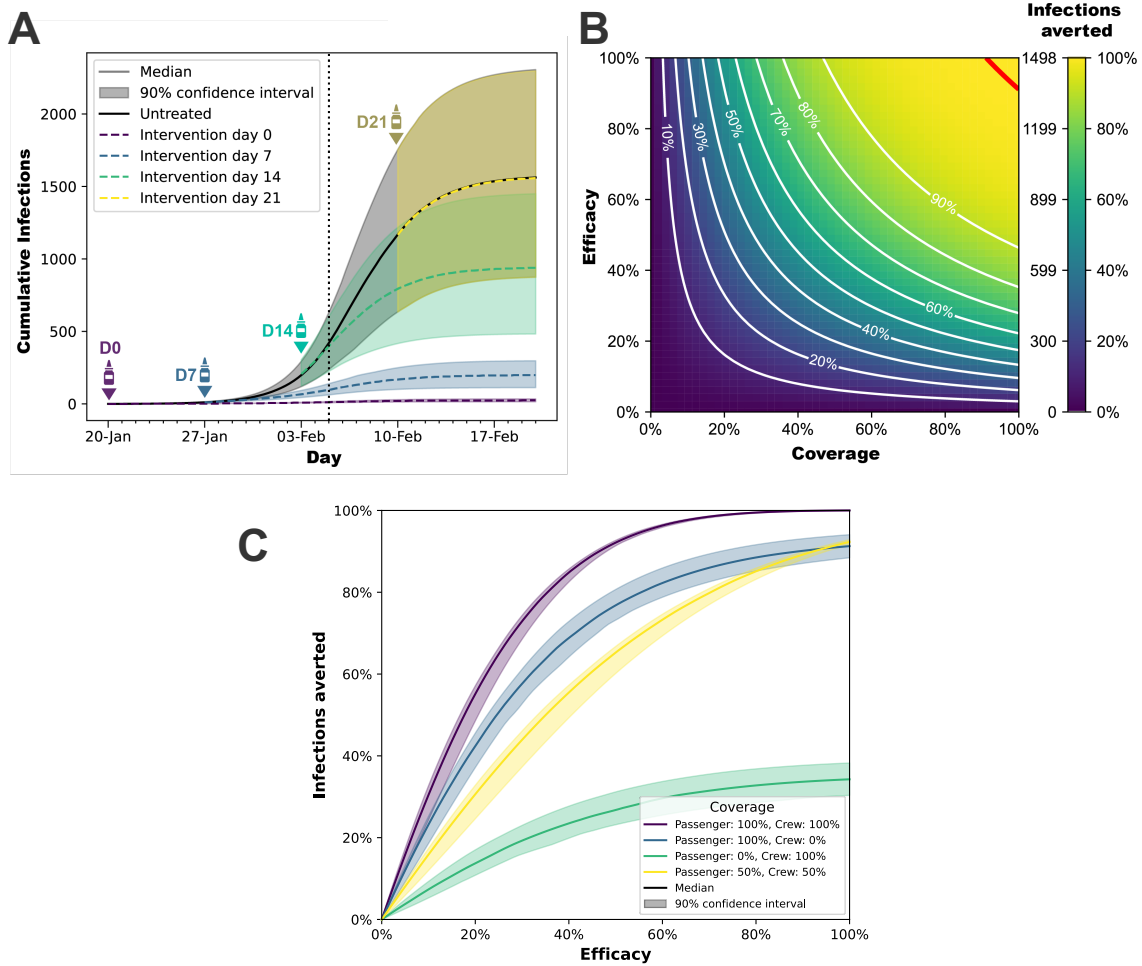

**Figure 3.** Impact of intranasal prophylaxis on the cruise ship. **A)** Number of cumulative infections over time for four scenarios in which an intranasal prophylaxis of 70% efficacy is administered to 100% of passengers and crew at day 0, day 7, day 14, or day 21 after first symptoms were detected. **B)** Sensitivity of the "mean percentage and number of infections averted" to the mean efficacy and the coverage of each treatment. The red line indicates the mean percentage of infections that need to be averted in order to reduce $R_0$ <1. **C)** Percentage of infections averted based on the efficacy of intranasal prophylaxis in four scenarios with different coverage among passengers and crew: 1) 100% coverage among passengers and crew, 2) 100% passengers only, 3) 100% of crew only, and 4) 50% coverage among passengers and 50% coverage among crew.

*Unknown superspreading event*

The second case study is the hypothetical outbreak of SARS-CoV-2 in a 5-day conference where data was collected on pairwise contacts over the duration of the conference. We consider the recorded interactions between individuals via a real-world weighted contact network (Figures 4A-C)[16]. Here, there is no testing or public health intervention, and the outbreak is unknown to the conference attendees. We assume that individuals take the nasal spray treatment as a preventative measure at their own discretion.

The conference contact network shows a high degree of connectivity, thus identifying several individuals as potential "superspreaders" with the ability to infect a significant number of network contacts (Figure 4A-C). While a significant proportion of these interactions are low frequency, leading to a low probability of infection, a small proportion of conference attendees interact over 1000 times, corresponding to a very high certainty of transmission. Mapping the frequency of pairwise interactions to contact network weightings results in a J-shaped distribution (long right-hand tail, with variance much bigger than the mean number of contacts), the details of which are discussed in the Supporting Information.

Because the incubation period for SARS-CoV-2 occurs on a similar timescale as the length of the conference itself, we see that the total number of infections in both treated and untreated scenarios peak shortly after the conference ends, around 7 days after the first infected cases (Figure 4E). Given that interactions between conference participants end on day 5, virus transmission ceases, and the number of daily infections is driven down.

We modelled the effect of suboptimal treatment adherence. Claxton et al. (2001) studied self-administration of medication in patients and found that an increase in the required dosing frequency led to an average decrease in treatment adherence[18,19] (Figure 4F). We assume that for each dosing scenario, the frequency of dosing is necessary to achieve a constant efficacy of 70% to make a fair comparison between each dosing scenario. In doing so, we decouple dosing regimen from efficacy, and focus only on the decrease in treatment adherence as dosing regimen increases. Suboptimal treatment adherence contributes to a decrease in the mean percentage of mild and severe infections averted (Figure 4G-H). However, the stochastic nature of these models leads to significant variability in outcomes, and thus a significant overlap in confidence intervals.

The simulations suggest that in the once daily dosing category where treatment adherence is highest (70-90%), prophylactic administration of a 70% efficacious treatment can only prevent ~30% of the mean number of mild infections. Although the simulations predict ~70% of severe infections would be prevented by this treatment, these outcomes are significantly worse than the number of infections averted for the same efficacy and coverage predicted in the Diamond Cruise ship case study. The inherent randomness and stochasticity in real-life contact patterns means that while some individuals interact frequently and with many people, there are also many individuals that have very few, infrequent pairwise interactions. This provides opportunities for infection to escape treatment via highly connected network nodes, inevitably reducing the treatment's population-level impact and increasing the variability in model predictions. Additionally, in the absence of supplementary social distancing or isolation of infected individuals, prophylactic treatments would certainly require much higher efficacies and coverages to achieve comparable reductions in infections.

However, use of a prophylactic treatment in this scenario still has an individual-level benefit. Across all dosing regimens considered in Figure 5, the percentage of individuals infected that take the prophylactic treatment is systematically less than in unprotected individuals. On average, an individual can almost half their probability of infection by receiving a treatment of 70% efficacy, significantly increasing an individual's chances of avoiding infection. Additionally, adhering to the prophylactic treatment also reduces an individual's probability of developing a severe infection as per the modelling assumptions outlined in the Methods Section. It also decreases the potential for fatality and long-term morbidity, alleviating an increased burden on healthcare systems.

**Table 2.** Summary of model parameters for Scenario II (SPHH Conference).

| PARAMETER | DESCRIPTION | VALUE |
|---|---|---|
| **INFECTION** | | |
| $N$ | Total number of participants | 405 [16] |
| $R_0$ | Basic reproduction number | 2.8 [20] |
| $a$ | Rate of infection | 1/3.4 days$^{-1}$ [20] |
| $p_s$ | Probability of developing a severe infection, given exposure | 0.45 [20] |
| $p_m = 1 - p_s$ | Probability of developing a mild infection, given exposure | 0.55 |
| $\gamma_s$ | Rate of recovery from severe infection | 1/5.7 days$^{-1}$ [20] |
| $\gamma_m$ | Rate of recovery from mild infection | 1/2.9 days$^{-1}$ [20] |
| $p_{p,s} = (1-\varepsilon)p_s$ | Probability of developing a severe infection, given exposure and partial protection | 0.135 |
| $p_{p,m} = 1 - p_{p,s}$ | Probability of developing a mild infection, given exposure and partial protection | 0.865 |
| $I_{s0}$ | Initial number of severe infections | 1 |
| $I_{m0}$ | Initial number of mild infections | 1 |
| **TREATMENT** | | |
| $\varepsilon$ | Treatment efficacy | 70% |
| $q$ | Population adherence to treatment | 79% ± 14% (once daily) [18,19] |

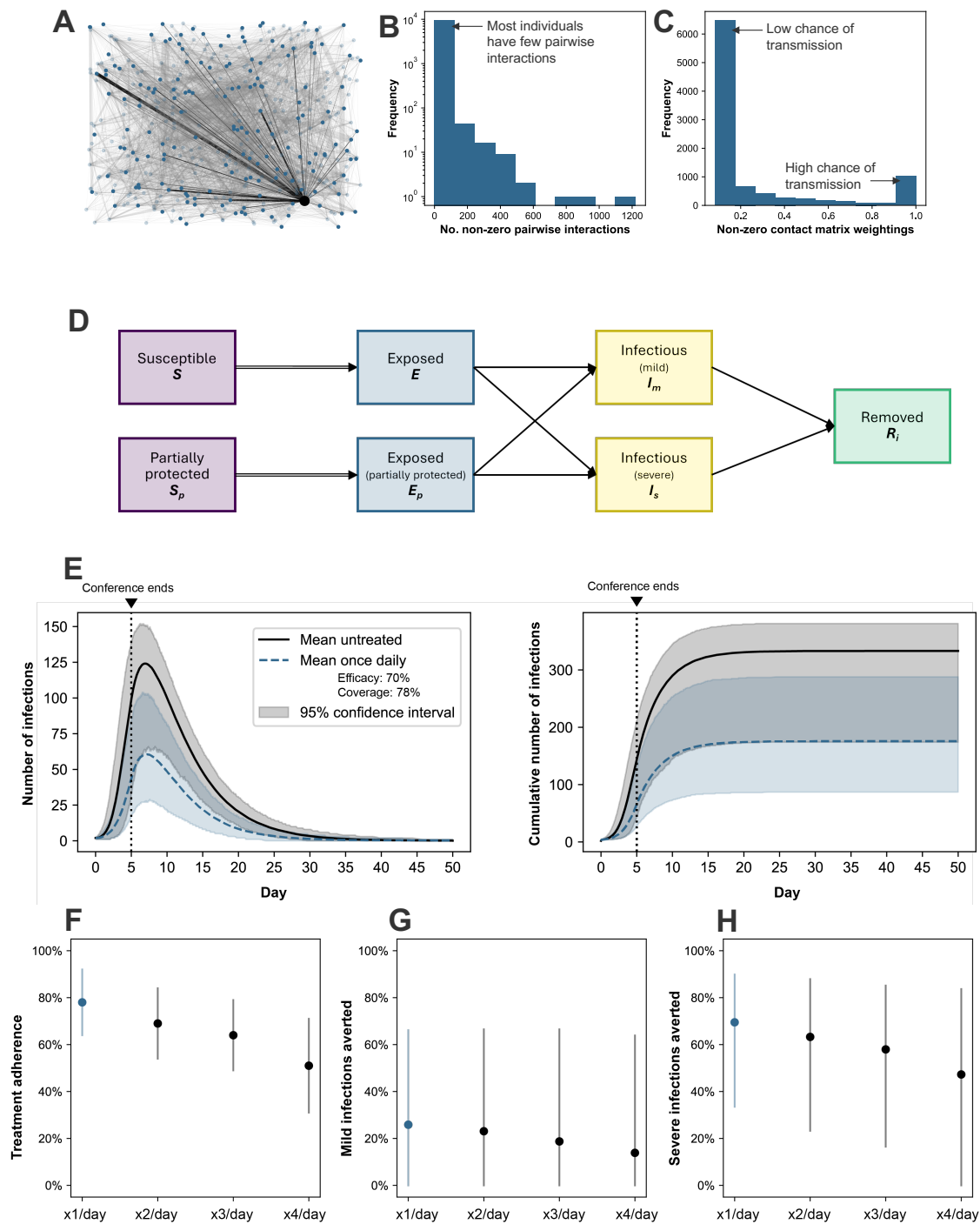

**Figure 4. A)** Undirected network schematic of the interactions between individuals in the conference scenario. **B)** Frequency of non-zero pairwise interactions between individuals. Most individuals interact with others for a small number of times. However, some individuals interact hundreds of times. **C)** Non-zero contact matrix weightings. Contact matrix weightings are determined by the frequency of pairwise interactions between individuals. It is assumed that the top 10% of interactions result in a weighting of 1, and the bottom 90% of interactions have a weighting that scales linearly between 0 and 1. **D)** Schematic of the SEIR model. All attendees

(apart from one mildly and one severely infected individual) begin the conference either susceptible or partially protected. Individuals can then be exposed to the virus, leading to a mild or serious infection, before recovering. **E)** Number of infections and cumulative number of infections over time. The solid black line denotes the mean number of infections assuming no prophylactic treatment administration, and the dotted blue line denotes the mean number of infections assuming prophylactic treatment that must be administered once daily. Shaded regions denote the corresponding 95% confidence interval. **F)** Percentage treatment adherence as a function of dosing regimen, reproduced from Claxton et al.[18,19] Black dots denote the mean value, with black lines representing ± 1 standard deviation. **G)** Percentage of mild infections averted as a function of dosing regimen. Black dots denote the mean value, with black lines representing the 95% confidence interval. **H)** Percentage of serious infections averted as a function of dosing regimen. Black dots denote the mean value, with black lines representing the 95% confidence interval.

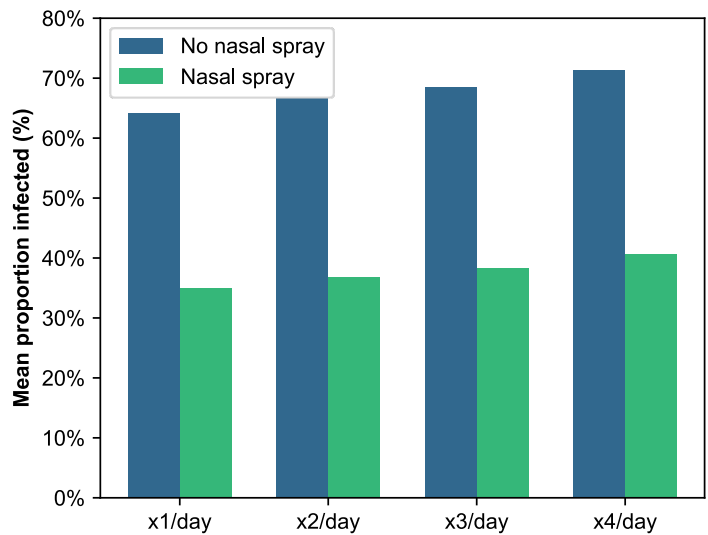

**Figure 5.** Mean proportion of individuals that are infected whilst having taken the intranasal spray (green) or not (blue). Each compliance scenario is based on dosing frequency (see Figure 4F) in the conference case study, assuming a constant 70% efficacy.

**Discussion**

The analyses of two case studies show how an intranasal spray can contribute to preventing superspreading events. During the known virus outbreak on the Diamond Princess Cruise Ship in 2020, intranasal sprays could have significantly reduced the effective reproductive number of the infection ($R$) and averted most of the infections, even if spray efficacy and target population coverage do not exceed 50%. During a short conference event with superspreading characteristics ('crowds, closed spaces and close contact') intranasal sprays can reduce individual attendees' risk of infection, but not the elimination of the virus from the event altogether and potential onward transmission when the event ends. This is because the conference duration is too short relative to the incubation period of the virus for treatment to have a major impact.

*Known outbreak in superspreading context*

During a known outbreak in a known superspreading context, intranasal sprays can be highly effective even at suboptimal efficacy or coverage. With efficacy and coverage over 65%, around 90% of infections can be averted. However, it is necessary to intervene within the first 1-2 weeks after the first case is identified. The earlier the intervention is rolled out, the greater the impact on infections averted.

Aboard the Diamond Princess cruise ship, people where quarantined, and infected people were removed from the population. These measures likely reduce the effective reproductive number of the virus compared to a superspreading event where free movement and contact is still possible. As cruise ship passengers and crew were allowed to disembark the ship only after infection, usage of the intranasal spray is both beneficial from a public health perspective and from an individual perspective.

The scenario of 100% intranasal spray coverage is similar to ring vaccination programmes used in Ebola and smallpox control[22–24]. Ring vaccination refers to vaccinating people who are socially or geographically connected to a known case. Our study suggests that such a campaign could also be effective against SARS-CoV-2 if using intranasal sprays, even if not all contacts of a closed population are treated. In contrast to ring vaccination, 'ring nose sprays' may also provide better protection also to elderly and immunocompromised, who typically have poorer immune response to vaccination. Ring vaccination has the advantage of durable protection, whereas intranasal sprays using monoclonal antibodies would provide protection only for as long as the active substance is at the site of infection since the systemic absorption is very limited[12].

Closed populations such as a cruise ship may be a useful setting for a field trial of a prophylactic intranasal spray against respiratory viruses. Some intranasal sprays have already shown efficacy in humans in field trials, but the event rates in closed populations tend to be higher, therefore, it is easier to establish proof-of-concept in such a setting.

*Event with potential superspreading characteristics*

In an event with potential superspreading characteristics like a conference, there is a higher level of uncertainty. Maybe one of the participants is infected with SARS-CoV-2, maybe nobody is. Compared to the cruise ship scenario, there is much less time at a five-day conference to identify an infected participant. Per case averted, many people would have to take the intranasal spray. For systematic prevention of superspreading of respiratory viruses at conferences in a pandemic scenario, all conferences would have to be targeted, even though not all of them would necessarily become superspreading events anyway. Identifying cases at the start of a conference with lateral flow testing could be a solution, although once a case is identified, it would be more effective if they were discouraged from attending the conference events at all. Nasal sprays therefore provide a possible solution to supplement testing, by providing protection against infection from individuals that escape positive identification via testing.

In conclusion, at a short event with crowds in closed spaces and close contacts, usage of intranasal sprays confers more of a benefit in protecting the individual user than from preventing community-level incidence.

*Sensitivity to assumptions*

It is important to note that conclusions drawn from this analysis must be considered subject to the underlying model assumptions. For example, we consider treatment coverage and efficacy to remain constant for the duration of each event. While this may be a fair assumption for short durations of time, such as the 5-day conference, this may fail to be true for longer timescales where treatment adherence varies over time as individuals lose interest or forget to take the treatment. Additionally, we neglect pre-existing immunity in each scenario, assuming that immunity is conferred via intranasal prophylaxis only. While capturing these details would provide more accurate predictions of virus transmission, detailed human behavioural data and estimates of population immunity would be required for analyses.

However, the simulations presented in this paper are not unique to SARS-CoV-2. In particular, the simulation of virus spread and prophylactic intervention presented in the SPHH conference scenario can be used to model a wide range of respiratory viruses and intervention strategies. For example, by varying the basic reproduction number and incubation period for a given virus strain, it is possible to tune the stochastic model to simulate the spread of many respiratory viruses in a real-life contact network. Additionally, the efficacy parameter in this model can be selected to reflect other interventions more accurately such as vaccination, social distancing, mask-wearing etc. This analysis provides a flexible framework for epidemiological modelling of a hypothetical superspreading event, appropriately capturing the probabilistic nature of chance in virus transmission in the real world.

**Conclusion**

In conclusion, our results suggest that at a known superspreading event, intranasal sprays can complement quarantining measures and potentially rapidly stop a SARS-CoV-2 outbreak, even at suboptimal coverage of the nasal spray. In contrast, at a superspreading event of short duration, intranasal sprays can reduce individuals' risk of infection, but cannot prevent all infections or onwards community transmission.

To maximize the impact of intranasal sprays in preventing superspreading events, the availability of fast and cheap diagnostics is key. Rapid diagnosis can turn a situation from 'an event with potential superspreading characteristics', where many people have to be treated to prevent a large percentage of the onward infections, to 'a closed population with known infections', where an intranasal spray can potentially extinguish the local outbreak.

**Author Contributions:** All authors contributed to study conception. The first draft of the manuscript was written by GB and KLR with input from all other authors, and later drafts of the manuscript were written by GB, AB, KLR and JF. Model development was done by GB, CH, RMA and FdW, and analyses were performed by GB. All authors contributed to data interpretation. All authors reviewed and commented on early versions of the manuscript, and all authors read and approved the final manuscript.

**Competing Interest Statement:** CH, RMA and FdW have no relevant financial or non-financial interests to disclose. GB, AB, ZM, KLR, JF and JG have disclosed that they are or were employees of Leyden Laboratories with stock options in this company. Leyden Laboratories develops intranasal technologies to administer substances for prevention and treatment of respiratory infections and non-communicable diseases, and is currently not involved in vaccine development, production or distribution.

**References**

1. F. Wong, J. J. Collins, Evidence that coronavirus superspreading is fat-tailed. *PNAS* **117**, 29416–29418 (2020).
2. D. C. Adam, *et al.*, Clustering and superspreading potential of SARS-CoV-2 infections in Hong Kong. *Nat Med* **26**, 1714–1719 (2020).
3. J. E. Lemieux, *et al.*, Phylogenetic analysis of SARS-CoV-2 in Boston highlights the impact of superspreading events. *Science (1979)* **371** (2021).
4. N. F. Che Mat, H. A. Edinur, M. K. A. Abdul Razab, S. Safuan, A single mass gathering resulted in massive transmission of COVID-19 infections in Malaysia with further international spread. *J Travel Med* **27** (2020).
5. A. L. Beukenhorst, *et al.*, SARS-CoV-2 elicits non-sterilizing immunity and evades vaccine-induced immunity: implications for future vaccination strategies. *Eur J Epidemiol* **38**, 237–242 (2023).
6. J. A. Backer, *et al.*, The impact of influenza vaccination on infection, hospitalisation and mortality in the Netherlands between 2003 and 2015. *Epidemics* **26**, 77–85 (2019).
7. A. Pan, *et al.*, Association of public health interventions with the epidemiology of the COVID-19 outbreak in Wuhan, China. *JAMA* **323**, 1915–1923 (2020).
8. C. Murphy, *et al.*, Effectiveness of social distancing measures and lockdowns for reducing transmission of COVID-19 in non-healthcare, community-based settings. *Philosophical Transactions of the Royal Society A* **381**, 20230132 (2023).
9. R. Song, *et al.*, Post-exposure prophylaxis with SA58 (anti-SARS-COV-2 monoclonal antibody) nasal spray for the prevention of symptomatic COVID-19 in healthy adult workers: a randomized, single-blind, placebo-controlled clinical study*. *Emerg Microbes Infect* **12** (2023).
10. X. Li, *et al.*, Real-world effectiveness of an intranasal spray A8G6 antibody cocktail in the post-exposure prophylaxis of COVID-19. *Signal Transduct Target Ther* **8**, 403 (2023).
11. Y. Liu, *et al.*, Human monoclonal antibody F61 nasal spray effectively protected high-risk populations from SARS-CoV-2 variants during the COVID-19 pandemic from late 2022 to early 2023 in China. *Emerg Microbes Infect* **13** (2024).
12. X. Zhang, *et al.*, Prophylactic efficacy of an intranasal spray with 2 synergetic antibodies neutralizing Omicron. *JCI Insight* **9** (2024).
13. C. Hadjichrysanthou, *et al.*, Exploring the Role of Antiviral Nasal Sprays in the Control of Emerging Respiratory Infections in the Community. *Infect Dis Ther* **11**, 2287–2296 (2022).
14. J. C. Emery, *et al.*, The contribution of asymptomatic SARS-CoV-2 infections to transmission on the Diamond Princess cruise ship. *Elife* **9**, e58699 (2020).
15. J. Stehlé, *et al.*, Simulation of an SEIR infectious disease model on the dynamic contact network of conference attendees. *BMC Med* **9**, 1–15 (2011).

16. M. Génois, A. Barrat, Can co-location be used as a proxy for face-to-face contacts? *EPJ Data Sci* **7**, 1–18 (2018).
17. D. T. Gillespie, Exact stochastic simulation of coupled chemical reactions. *J Phys Chem* **81**, 2340–2361 (1977).
18. L. Osterberg, T. Blaschke, Adherence to medication. *New England Journal of Medicine* **353**, 487–497 (2005).
19. A. J. Claxton, J. Cramer, C. Pierce, A systematic review of the associations between dose regimens and medication compliance. *Clin Ther* **23**, 1296–1310 (2001).
20. P. Zhu, V. Zhang, A. L. Wagner, Demographic differences in compliance with COVID-19 vaccination timing and completion guidelines in the United States. *Vaccines (Basel)* **11**, 369 (2023).
21. S. Merler, *et al.*, Containing Ebola at the source with ring vaccination. *PLoS Negl Trop Dis* **10**, e0005093 (2016).
22. M. Kretzschmar, S. Van den Hof, J. Wallinga, J. Van Wijngaarden, Ring vaccination and smallpox control. *Emerg Infect Dis* **10**, 832 (2004).
23. E. ça S. R. V. T. Consortium, The ring vaccination trial: a novel cluster randomised controlled trial design to evaluate vaccine efficacy and effectiveness during outbreaks, with special reference to Ebola. *BMJ: British Medical Journal* **351**, h3740 (2015).
24. R. H. Johnstone, *et al.*, Uncertainty and variability in models of the cardiac action potential: Can we build trustworthy models? *J Mol Cell Cardiol* **96**, 49–62 (2016).
25. M. Clerx, *et al.*, Probabilistic Inference on Noisy Time Series (PINTS). *J Open Res Softw* **7**, 23 (2019).

**Supporting Information**

**Governing equations for Scenario I: Diamond Princess cruise ship**

We model the transmission of SARS-CoV-2 aboard the Diamond Princess cruise ship via an adapted SEIR model from Emery et al. (2020)[1]

$$\frac{dS_p}{dt} = -\beta_{pp}\left(\frac{\theta_a A_p + \theta_p P_p + I_{skp} + I_{sup}}{N_p}\right)S_p \\ - \beta_{pc}\left(\frac{\theta_a A_c + \theta_p P_c + I_{skc} + I_{suc}}{N_c}\right)S_p, \quad \text{(S1a)}$$

$$\frac{dS_c}{dt} = -\beta_{cp}\left(\frac{\theta_a A_p + \theta_p P_p + I_{skp} + I_{sup}}{N_p}\right)S_c - \beta_{cc}\left(\frac{\theta_a A_c + \theta_p P_c + I_{skc} + I_{suc}}{N_c}\right)S_c, \quad \text{(S1b)}$$

$$\frac{dS_{p,p}}{dt} = -\beta_{pp}(1-\varepsilon)\left(\frac{\theta_a A_p + \theta_p P_p + I_{skp} + I_{sup}}{N_p}\right)S_{p,p} \\ -\beta_{pc}(1-\varepsilon)\left(\frac{\theta_a A_c + \theta_p P_c + I_{skc} + I_{suc}}{N_c}\right)S_{p,p}, \quad \text{(S1c)}$$

$$\frac{dS_{p,c}}{dt} = -\beta_{cp}(1-\varepsilon)\left(\frac{\theta_a A_p + \theta_p P_p + I_{skp} + I_{sup}}{N_p}\right)S_{p,c} \\ -\beta_{cc}(1-\varepsilon)\left(\frac{\theta_a A_c + \theta_p P_c + I_{skc} + I_{suc}}{N_c}\right)S_{p,c}, \quad \text{(S1d)}$$

$$\frac{dE_p}{dt} = \beta_{pp}\left(\frac{\theta_a A_p + \theta_p P_p + I_{skp} + I_{sup}}{N_p}\right)(S_p + (1-\varepsilon)S_{p,p}) \\ + \beta_{pc}\left(\frac{\theta_a A_c + \theta_p P_c + I_{skc} + I_{suc}}{N_c}\right)(S_p + (1-\varepsilon)S_{p,p}) - \nu E_p, \quad \text{(S1e)}$$

$$\frac{dE_c}{dt} = \beta_{cp}\left(\frac{\theta_a A_p + \theta_p P_p + I_{skp} + I_{sup}}{N_p}\right)(S_c + (1-\varepsilon)S_{p,c}) \\ - \beta_{cc}\left(\frac{\theta_a A_c + \theta_p P_c + I_{skc} + I_{suc}}{N_c}\right)(S_c + (1-\varepsilon)S_{p,c}) - \nu E_c, \quad \text{(S1f)}$$

$$\frac{dF_p}{dt} = \nu(E_p - F_p), \quad \text{(S1g)}$$

$$\frac{dF_c}{dt} = \nu(E_c - F_c), \quad \text{(S1h)}$$

$$\frac{dP_p}{dt} = (1-\chi)\nu F_p - (\phi + \gamma_p)P_p, \quad \text{(S1i)}$$

$$\frac{dP_c}{dt} = (1-\chi)\nu F_c - (\phi + \gamma_p)P_c, \quad \text{(S1j)}$$

$$\frac{dA_p}{dt} = \chi\nu F_p - (\phi + \gamma_a)A_p, \quad \text{(S1k)}$$

$$\frac{dA_c}{dt} = \chi\nu F_c - (\phi + \gamma_a)A_c, \quad \text{(S1l)}$$

$$\frac{dI_{skp}}{dt} = f\gamma_p P_p - (m + \gamma_s)I_{skp}, \quad \text{(S1m)}$$

$$\frac{dI_{skc}}{dt} = f\gamma_p P_c - (m + \gamma_s)I_{skc}, \quad \text{(S1n)}$$

$$\frac{dI_{sup}}{dt} = (1-f)\gamma_p P_p - (m + \gamma_s)I_{sup}, \quad \text{(S1o)}$$

$$\frac{dI_{suc}}{dt} = (1-f)\gamma_p P_c - (m + \gamma_s)I_{suc}, \quad \text{(S1p)}$$

$$\frac{dR_p}{dt} = m(I_{skp} + I_{sup}) + \phi(P_p + A_p + R_{+p}), \quad \text{(S1q)}$$

$$\frac{dR_c}{dt} = m(I_{skc} + I_{suc}) + \phi(P_c + A_c + R_{+c}), \tag{S1r}$$

$$\frac{dR_{+p}}{dt} = \gamma_s(I_{skp} + I_{sup}) + \gamma_a A_p - (h + \phi)R_{+p}, \tag{S1s}$$

$$\frac{dR_{+c}}{dt} = \gamma_s(I_{skc} + I_{suc}) + \gamma_a A_c - (h + \phi)R_{+c}, \tag{S1t}$$

$$\frac{dR_{-p}}{dt} = hR_{+p}, \tag{S1u}$$

$$\frac{dR_{-c}}{dt} = hR_{+c}, \tag{S1v}$$

where

$$\beta_{ij}(t) = (\bar{\beta} + a_2)c_{ij}\left(1 - \frac{b_1}{1 + \exp\left(-b_2(t - \tau_{ij} - a_1)\right)}\right), \tag{S2}$$

for $i,j \in \{p,c\}$. Definitions for individual compartments are described in Figure 1. Equation S2 captures temporal variations in the transmission rate between passengers and crew via a logistic function, with parameters summarised in Table 1. Equations S1 are solved subject to the initial conditions

$$S_p = (1 - q_p)N_p - 1, \quad S_c = (1 - q_c)N_c, \quad S_{p,p} = q_p N_p, \quad S_{p,c} = q_c N_c,$$
$$I_{skp} = 1 \quad \text{at} \quad t = 0, \tag{S3}$$

with the initial condition for remaining compartments equal to 0.

To parameterise our model using the Diamond Princess passenger and crew infection-time data, we adapt the fitting procedure approach from Emery et al. (2020) for the same scenario[1]. That is, a proportion of parameters are fixed as per literature estimates, as summarised in Table 1, with the remaining parameters fitted to data via Bayesian inference methods. For fitted parameters, we prescribe uniform log prior distribution estimates (Table S1) and assume a Gaussian log likelihood. In doing so, we introduce two parameters, $\sigma_p$ and $\sigma_c$, which describe the standard deviation in passenger and crew infection measurements, respectively. Posterior distributions are sampled using the Haario Bardenet Adaptive Covariance MCMC algorithm implemented in the open-source Python package, PINTS[2]. Posterior samples are run for 100,000 iterations with 5000 initial adaptation-free iterations, and we discard the first 20000 warm-up iterations. The resulting marginal and pairwise posterior distributions can be seen in Figure S1. Table S2 provides a summary of statistics quantifying the goodness of fit including Highest Density Interval (HDI), effective sample size (ESS), and $\hat{R}$ scores for each parameter.

**Deterministic model for Scenario II: SPHH Conference**

We model the transmission of SARS-CoV-2 on an undirected graph describing interactions of individuals at a conference via a stochastic adaptation of the classic SEIR model. The analogous deterministic model in a homogeneous population is given by the following ordinary differential equations

$$\frac{dS}{dt} = -\beta\left(\frac{I_m + I_s}{N}\right)S, \tag{S4a}$$

$$\frac{dS_p}{dt} = -(1 - \varepsilon)\beta\left(\frac{I_m + I_s}{N}\right)S_p, \tag{S4b}$$

$$\frac{dE}{dt} = \beta\left(\frac{I_m + I_s}{N}\right)S - aE, \tag{S4c}$$

$$\frac{dE_p}{dt} = (1 - \varepsilon)\beta\left(\frac{I_m + I_s}{N}\right)S_p - aE_p, \tag{S4d}$$

$$\frac{dI_m}{dt} = ap_m E + ap_{p,m} E_p - \gamma_m I_m, \tag{S4e}$$

$$\frac{dI_s}{dt} = ap_s E + ap_{p,s} E_p - \gamma_s I_s, \tag{S4f}$$

$$\frac{dR}{dt} = \gamma_m I_m + \gamma_s I_s, \tag{S4g}$$

subject to the initial conditions,

$$S = S_0, \ S_p = 0, \ E = 0, \ E_p = 0, \ I_m = I_{m0}, \ I_s = I_{s0}, \ R = 0 \quad \text{at} \quad t = 0. \tag{S5}$$

We note that in S4, we assume that the transmission rate $\beta$ is constant and equal for infections from both mild and severe individuals.

To determine the transmission rate $\beta$ from a given basic reproduction number $R_0$, we consider the disease-free equilibrium prior to an outbreak, where prophylactic treatment has not been administered. In this scenario, the number of susceptible individuals is simply equal to the total population, and the number of partially protected susceptible individuals is zero:

$$S^* = N, \ S_p^* = 0, \tag{S6}$$

where superscript asterisks denote the disease-free equilibrium value in each compartment. The basic reproduction number is the dominant eigenvalue of the next generation matrix, $FV^{-1}$. For S4, the matrices $F$ and $V$ are defined as

$$F = \begin{bmatrix} 0 & \beta & \beta \\ 0 & 0 & 0 \\ 0 & 0 & 0 \end{bmatrix}, \quad V = \begin{bmatrix} a & 0 & 0 \\ -ap & \gamma_m & 0 \\ -a(1-p) & 0 & \gamma_s \end{bmatrix}. \tag{S7}$$

Therefore, the inverse of $V$ is given to be

$$V^{-1} = \begin{bmatrix} \dfrac{1}{a} & 0 & 0 \\ \dfrac{p}{\gamma_m} & \dfrac{1}{\gamma_m} & 0 \\ \dfrac{1-p}{\gamma_s} & 0 & \dfrac{1}{\gamma_s} \end{bmatrix}. \tag{S8}$$

Combining S7 and S8, the next generation matrix is given by

$$FV^{-1} = \begin{bmatrix} \dfrac{p\beta}{\gamma_m} + \dfrac{(1-p)\beta}{\gamma_s} & \dfrac{\beta}{\gamma_m} & \dfrac{\beta}{\gamma_s} \\ 0 & 0 & 0 \\ 0 & 0 & 0 \end{bmatrix}, \tag{S9}$$

with corresponding eigenvalues

$$\lambda_1 = \frac{\beta(p\gamma_s + (1-p)\gamma_m)}{\gamma_m \gamma_s}, \quad \lambda_{2,3} = 0. \tag{S10a,b}$$

Taking the dominant eigenvalue from S9, we can define the basic reproduction to be

$$R_0 = \frac{\beta(p\gamma_s + (1-p)\gamma_m)}{\gamma_m \gamma_s}. \tag{S11}$$

We use Equation S11 to relate a prescribed $R_0$ value to the infection rate $\beta$ parameterizing Equations S4.

**Stochastic model for Scenario II: SPHH Conference**

In this paper, we model the stochastic analogue of Equations S4, where individuals belong to a given state, and transitions between states are governed probabilistically. To initiate the stochastic simulation, we randomly infect two individuals - one with a mild infection, and one with a severe infection (analogous to S5). The model then stochastically simulates the spread of the infection over time, based on the interactions represented in the weighted contact network and the disease progression dynamics.

State transitions are modelled via the Gillespie algorithm[3] - a stochastic simulation technique that allows us to model the system's evolution where events occur continuously and randomly over time. At each step, the algorithm calculates the rates of all possible events (transitions between states and infection events) based on the current state of the system. It then randomly selects the next event to occur based on these rates, using a weighted random draw. The time until this event occurs is determined by sampling from an exponential distribution, ensuring that events occur continuously over time. When an event is selected, the system's state is updated accordingly. For example, if the selected event is the transition of an individual from Susceptible to Exposed, the individual's state is updated, and the overall counts of Susceptible and Exposed individuals are adjusted. This update also affects the event rates, which are recalculated at each step based on the new system state.

The contact network plays a critical role in determining the infection events. Individuals are more likely to transmit the virus to those they are connected to in the network. Our model accommodates a weighted contact matrix, reflecting the strength and frequency of interactions between individuals.

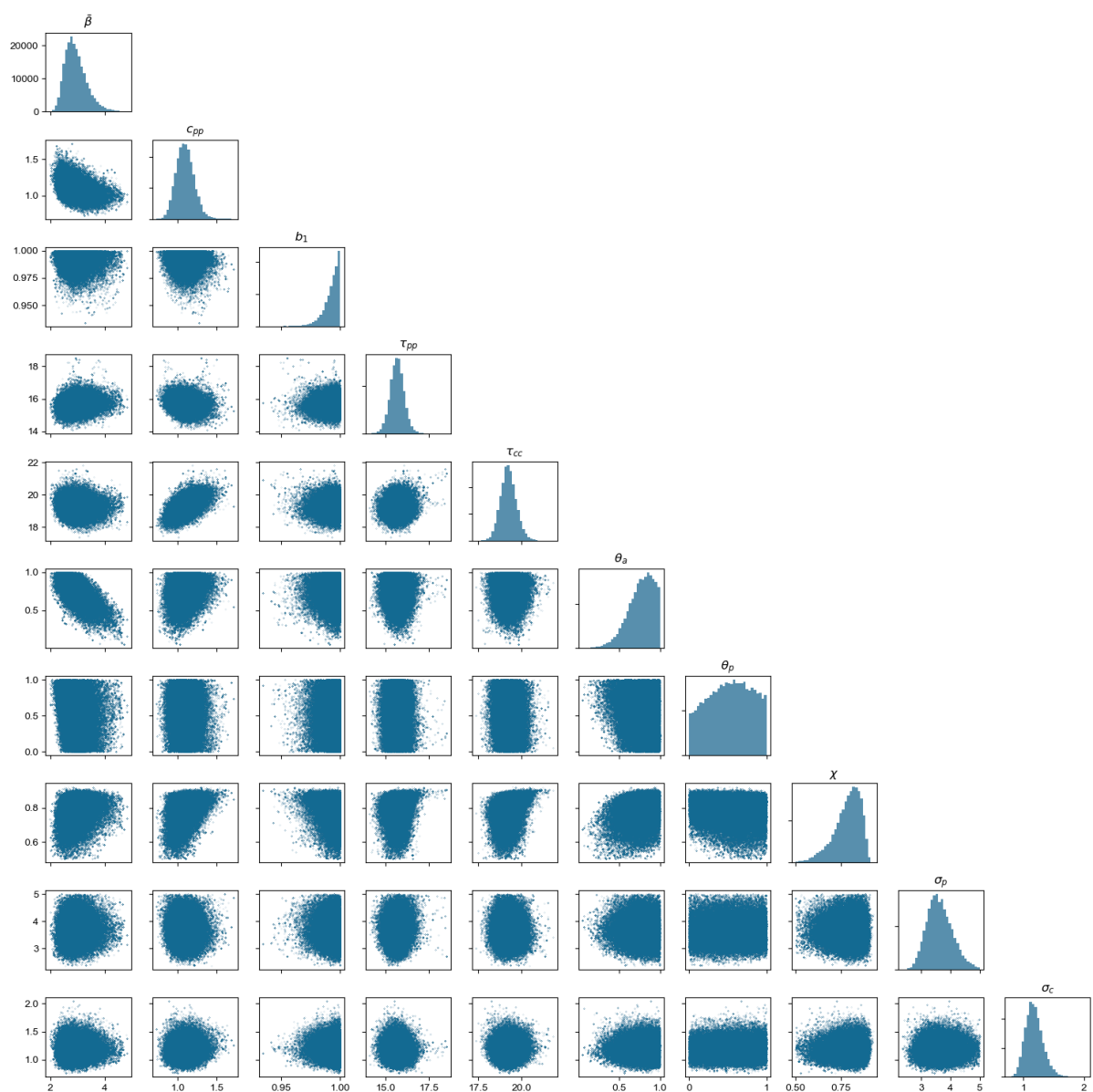

**Fig. S1.** Pair plot of the sampled posterior distributions for each fitted model parameter. Off-diagonal subplots denote the pairwise scatter plot for two model parameters. The leading diagonal denotes the individual posterior distributions.

**Table S1.** Summary of model priors for Scenario I (Diamond Princess).

| PARAMETER | PRIORS |
|---|---|
| $\bar{\beta}$ | Uniform(0, 6) |
| $c_{pp}$ | Uniform(0.5, 2) |
| $b_1$ | Uniform(0.8, 1) |
| $\tau_{pp} = \tau_{cp} = \tau_{pc}$ | Uniform(10, 20) |
| $\tau_{cc}$ | Uniform(15, 25) |
| $\theta_a$ | Uniform(0, 1) |
| $\theta_p$ | Uniform(0, 1) |
| $\chi$ | Uniform(0.5, 1) |
| $\sigma_p$ | Uniform(0, 5) |
| $\sigma_c$ | Uniform(0, 5) |

**Table S2.** Summary statistics for the model fitting of Scenario I (Diamond Princess), including the mean, standard deviation (SD), 94% Highest Density Interval (HDI), Markov Chain Standard Error (MCSE) of the mean and standard deviation, the bulk and tail Effective Sample Sizes (ESS), and the $\widehat{R}$ score, for each fitted parameter.

|  | Mean | SD | 94% HDI | MCSE Mean | MCSE SD | ESS Bulk | ESS Tail | $\widehat{R}$ |
|---|---|---|---|---|---|---|---|---|
| $\bar{\beta}$ | 2.851 | 0.364 | 2.221 - 3.546 | 0.015 | 0.011 | 607 | 1839 | 1.01 |
| $c_{pp}$ | 1.098 | 0.116 | 0.891 - 1.324 | 0.002 | 0.002 | 2296 | 4956 | 1.00 |
| $b_1$ | 0.993 | 0.006 | 0.981 - 1 | 0 | 0 | 2075 | 2066 | 1.00 |
| $\tau_{pp}$ | 15.648 | 0.38 | 14.946 - 16.375 | 0.006 | 0.005 | 3606 | 6225 | 1.00 |
| $\tau_{cc}$ | 19.231 | 0.458 | 18.344 - 20.063 | 0.009 | 0.006 | 2662 | 6082 | 1.00 |
| $\theta_a$ | 0.781 | 0.148 | 0.528 - 1 | 0.003 | 0.002 | 1934 | 3794 | 1.00 |
| $\theta_p$ | 0.546 | 0.272 | 0.101 - 1 | 0.006 | 0.004 | 2480 | 9293 | 1.01 |
| $\chi$ | 0.765 | 0.083 | 0.601 – 0.894 | 0.007 | 0.005 | 153 | 728 | 1.03 |
| $\sigma_p$ | 3.643 | 0.423 | 2.872 – 4.447 | 0.007 | 0.005 | 3809 | 7878 | 1.00 |
| $\sigma_c$ | 1.18 | 0.142 | 0.925 - 1.45 | 0.002 | 0.002 | 4193 | 6542 | 1.00 |

**Supporting Code**

Code used to generate figures can be found at https://github.com/gjibooth/ClosedPopulations.

**SI References**

1. J. C. Emery, *et al.*, The contribution of asymptomatic SARS-CoV-2 infections to transmission on the Diamond Princess cruise ship. *Elife* **9**, e58699 (2020).
2. M. Clerx, *et al.*, Probabilistic Inference on Noisy Time Series (PINTS). *J Open Res Softw* **7**, 23 (2019).
3. D. T. Gillespie, Exact stochastic simulation of coupled chemical reactions. *J Phys Chem* **81**, 2340–2361 (1977).